\documentclass[twocolumn,aps,preprintnumbers,amsmath,amssymb,superscriptaddress,floatfix]{revtex4-1}
\usepackage{subfigure,multirow}
\usepackage{longtable}

\usepackage{graphicx, color}
\usepackage{physics}
\usepackage{chngpage}

\usepackage{dcolumn}
\usepackage{bm}
\usepackage[T1]{fontenc}
\usepackage{lmodern}
\usepackage{xcolor}
\usepackage{makecell}

\makeatletter
\newcommand{\includegraphicsifexists}[2][]{%
  \IfFileExists{#2}{\includegraphics[#1]{#2}}{%
    \fbox{%
      \parbox[c][.55\linewidth][c]{.95\linewidth}{\centering \textbf{Figure placeholder:} \texttt{#2} not found in submission.}%
    }%
  }%
}
\makeatother
\usepackage{soul}

\begin{document}

\title{Revisiting Reactor Anti-Neutrino 5 MeV Bump with $^{13}$C Neutral-Current Interaction}

\author{Pouya Bakhti}
\email{pouya\_bakhti@jbnu.ac.kr}
\affiliation{Laboratory for Symmetry and Structure of the Universe, Department of Physics, Jeonbuk National University, Jeonju, Jeonbuk 54896, Korea}

\author{Min-Gwa Park}
\email{mgpark@jbnu.ac.kr}
\affiliation{Laboratory for Symmetry and Structure of the Universe, Department of Physics, Jeonbuk National University, Jeonju, Jeonbuk 54896, Korea}
\affiliation{Department of Physics and Astronomy, Northwestern University, Evanston, IL 60208, USA}

\author{Meshkat Rajaee}
\email{meshkat@jbnu.ac.kr}
\affiliation{Laboratory for Symmetry and Structure of the Universe, Department of Physics, Jeonbuk National University, Jeonju, Jeonbuk 54896, Korea}

\author{Chang Sub Shin}
\email{csshin@cnu.ac.kr}
\affiliation{Department of Physics and Institute of Quantum Systems, \\
Chungnam National University, Daejeon 34134, Korea}
\affiliation{Center for Theoretical Physics of the Universe, \\
Institute for Basic Science, Daejeon 34126, Korea}
\affiliation{Korea Institute for Advanced Study, Seoul 02455, Korea}

\author{Seodong Shin}
\email{sshin@jbnu.ac.kr}
\affiliation{Laboratory for Symmetry and Structure of the Universe, Department of Physics, Jeonbuk National University, Jeonju, Jeonbuk 54896, Korea}
\affiliation{Center for Theoretical Physics of the Universe, \\
Institute for Basic Science, Daejeon 34126, Korea}

\begin{abstract}

For the first time, we comprehensively examine the potential of a neutral-current interaction of reactor neutrino with $^{13}$C 
emitting a 3.685 MeV photon to identify the origin of the 5 MeV bump in reactor antineutrino spectra observed through the inverse beta decay (IBD) process. This anomaly may be due to new physics, reactor antineutrino flux inaccuracies, or IBD systematics. 
The 3.685 MeV photon released during the de-excitation of $^{13}$C$^\ast$ to its ground state is observable in 
liquid scintillator detectors. 
Remarkably, we confirm the powerfulness of our proposal by completely ruling out a new physics scenario explaining the bump from the existing NEOS data. 
We also explore the potential of current and forthcoming experiments, including solar neutrino studies at JUNO, pion and muon decay-at-rest experiments at OscSNS, and isotope decay-at-rest studies at Yemilab, to 
measure the cross-section precisely enough to distinguish the expected bump and the theoretical flux models via our channel.
Additionally, we propose a novel method to track the time evolution of reactor isotopes by analyzing the $^{13}$C signal, which 
yields critical insights into the contributions of $^{235}$U and $^{239}$Pu to the 
bump, acting as a robust tool. 

\end{abstract}

\maketitle

\section{Introduction}

Since its first discovery~\cite{Cowan:1956rrn}, reactor neutrinos have advanced our comprehension of the lepton sector. 
The KamLAND confirmed the neutrino oscillation as the explanation for the solar neutrino problem~\cite{KamLAND:2013rgu}.
Daya Bay~\cite{DayaBay:2018yms},  RENO~\cite{RENO:2018dro}, and  Double Chooz~\cite{DoubleChooz:2014kuw} measured the oscillation parameter $\theta_{13}$, which
has opened the possibility of CP violation in neutrino oscillation.
Future reactor experiment JUNO~\cite{JUNO:2015zny} aims to determine the neutrino mass ordering and achieve subpercent precision in measuring solar neutrino oscillation parameters, which can be complemented by the proposed liquid scintillator counter $\nu$EYE at Yemilab ($\nu$ Experiment at YEmilab)~\cite{Seo:2023xku, Bakhti:2023vzn}, promising
precise investigations into solar characteristics.

Probes of neutrino oscillation using reactor neutrinos necessitate accurate theoretical predictions of the neutrino flux. However, its calculation is very complicated and traces its origins to the early days subsequent to the first detection of neutrinos in the Cowan and Reines reactor experiment~\cite{Cowan:1956rrn}. 
The Vogel model~\cite{Vogel:1989iv}, employed as the standard flux model for two decades starting from the 1990s, relies on the conversion method and predicts reactor flux measurements based on the Institut Laue-Langevin (ILL) electron spectrum measurements~\cite{Schreckenbach:1981wlm, VonFeilitzsch:1982jw}.  

Interestingly, the actual neutrino fluxes measured in several short-baseline reactor neutrino experiments with varying fission fractions were smaller than the expectations,
which has drawn more careful calculations, like 
those by Mueller et al.~\cite{Mueller:2011nm} and Huber~\cite{Huber:2011wv}
called Huber-Mueller (HM) model.
Nonetheless, the predicted fluxes still surpass the observations by about 6\% corresponding to a $3 \sigma$ level discrepancy in the overall energy spectra, dubbed as the reactor antineutrino anomaly (RAA)~\cite{Mention:2011rk}.

Moving the focus to a narrower prompt energy range of 4 to 6 MeV, the situation becomes more arduous.
In contrast to the deficit in the overall flux, the observed data first reported by RENO~\cite{RENO:2015ksa}, 
and then confirmed by other experiments such as Daya Bay~\cite{DayaBay:2015lja}, Double Chooz~\cite{DoubleChooz:2014kuw}, NEOS~\cite{NEOS:2016wee}, Neutrino-4~\cite{Serebrov:2020kmd}, and DANSS (preliminary)~\cite{Danilov:2022bss} 
shows an excess
at more than 4$\sigma$.
Moreover, STEREO~\cite{STEREO:2020hup} and PROSPECT~\cite{PROSPECT:2022wlf}, utilizing research reactors powered by 100\% $^{235}$U fuel, have excluded the no-bump scenario with more than 3.5$\sigma$ and 2$\sigma$ C.L. respectively. This anomaly is called the {\it 5 MeV bump}~\cite{Huber:2016xis}.
Note that the bump gets exacerbated if the RAA is resolved by new flux models so that the overall flux is fitted to the observed spectrum.

Reactor electron anti-neutrinos have been
detected through the inverse beta decay (IBD) reaction, offering distinct advantages such as large cross-section, clear coincidence of the prompt positron annihilation and the delayed neutron capture $\gamma$-ray emission, and the low interaction threshold around 1.8 MeV. 
The 5 MeV bump means the energy distribution of the positron from IBD  peaks around 5 MeV beyond the theoretical expectations.
To further investigate its origin, whether it is coming from a miscalculation of the flux, new physics beyond the Standard Model (BSM), or IBD systematic, employing another detection approach is advantageous. 
In this paper, for the first time, we propose to use the neutral-current (NC) interaction of neutrinos with $^{13}$C isotope ubiquitous with mostly about 1.1\% abundance in the carbon-based  liquid scintillator~(LS) detectors to identify the 5 MeV bump alternatively to the IBD.
The interaction threshold is 3.685 MeV, resulting in the excitation of $^{13}$C nuclei to their first excited state. Then, a 3.685 MeV $\gamma$-ray emits from prompt de-excitation of $^{13}$C$^*$, creating a distinctive signal. 
We also investigate the possibility of reducing the cross-section uncertainties of this process from various experiments.
Note that NC interactions are flavor blind and are not affected by neutrino oscillation parameter uncertainties.
In contrast, other channels 
such as neutrino-electron elastic scattering (ES)
and coherent elastic neutrino-nucleus scattering (CE$\nu$NS) 
suffer from large systematic uncertainty and the background contamination, although their interaction threshold energies are low. 
However, note that CC and NC interactions of deuterons can be alternative channels in experiments with a sufficient quantity of heavy water. A review of detection channels for reactor antineutrinos is found in Ref.~\cite{Qian:2018wid}.

Comparisons of the measurements via IBD and our 3.685 MeV $\gamma$-ray channel can help identify the origin of 5 MeV bump: i) a miscalculation or modification (by BSM) of the flux if the bumps are comparable, ii) IBD systematics or the BSM contributions mimicking the IBD (but not our 3.685 MeV photon channel) if no-bump is observed in our channel, and iii) a BSM contribution, possibly containing light new particles, both mimicking the IBD and inducing a 3.685 MeV photon peak if the 5 MeV bump {\it via our channel} becomes more severe.
As an example of the last category, we examine the explanation of the 5 MeV bump by the new physics scenario involving a sterile neutrino with an additional U(1) gauge symmetry~\cite{Berryman:2018jxt}.
By directly searching for the 3.685 MeV photon signal much beyond the background events in the recent NEOS data, we conclusively rule out the vector interaction scenario, showing the effectiveness of our channel.

In the case that bump results from IBD systematics, measurements of the $^{13}$C flux are expected to align with flux models. As an example of IBD systematics, reconstructing prompt energy in liquid scintillators involves several complexities. The light yield demonstrates non-linearity with respect to the energy deposited by charged particles, primarily due to the quenching effect, as described by Birks' law
~\cite{Qian:2018wid,birks2013theory}, and contributions from Cherenkov radiation. Furthermore, electronic readout systems may introduce additional non-linearity by inaccurately estimating the number of detected photoelectrons. Variations in detector response, depending on the event position, further exacerbate these challenges. Consequently, comprehensive calibration procedures are critical for accurate prompt energy reconstruction. To address the 5-MeV bump effectively and verify that it is not a systematic artifact, alternative methods, such as analyzing $^{13}$C events, are indispensable.

To assess the broader feasibility of our proposed channel beyond the specific scenario discussed, we perform a detailed analysis of background events and calculate the signal-to-background ratio required to differentiate between various reactor neutrino flux models. 
Moreover, the most significant systematic uncertainty in differentiating reactor flux models arises from cross-section measurements. To address this issue, we explore the potential of current and future experiments—such as solar neutrino studies at JUNO, pion and muon decay-at-rest experiments at OscSNS, and isotope decay-at-rest studies at Yemilab—to achieve precise and accurate cross-section measurements. 

The structure of the paper is as follows. In Sec.~\ref{sec:general}, we discuss the expected number of events of our channel $\bar \nu_e ^{13}{\rm C} \to \bar \nu_e {}^{13}{\rm C}^\ast$ in the energy region of 4--8 MeV where the 5 MeV bump is situated. Section~\ref{sec:NP} examines the aforementioned BSM scenario that addresses the 5 MeV bump using $^{13}$C events at NEOS. In Sec.~\ref{sec:xs}, we provide a detailed analysis of the $^{13}$C cross-section measurement. Section~\ref{sec:fluxsens} investigates the potential for differentiating reactor flux models and the expected 5 MeV bump as well as the application in detecting fission fraction evolution. Finally, our conclusions are presented in Sec.~\ref{sec:conclusions}.
The detailed explanations on how the BSM scenario in Ref.~\cite{Berryman:2018jxt} contributes to our channel are in Appendix~\ref{appendix:nus} and the proposals of more background mitigation strategy in near-site reactor experiments are summarized in Appendix~\ref{appendix:moresubtraction}.

\section{Expected event spectrum}
\label{sec:general}

\begin{figure}[tbp]
\begin{centering}
\includegraphics[width=\linewidth]{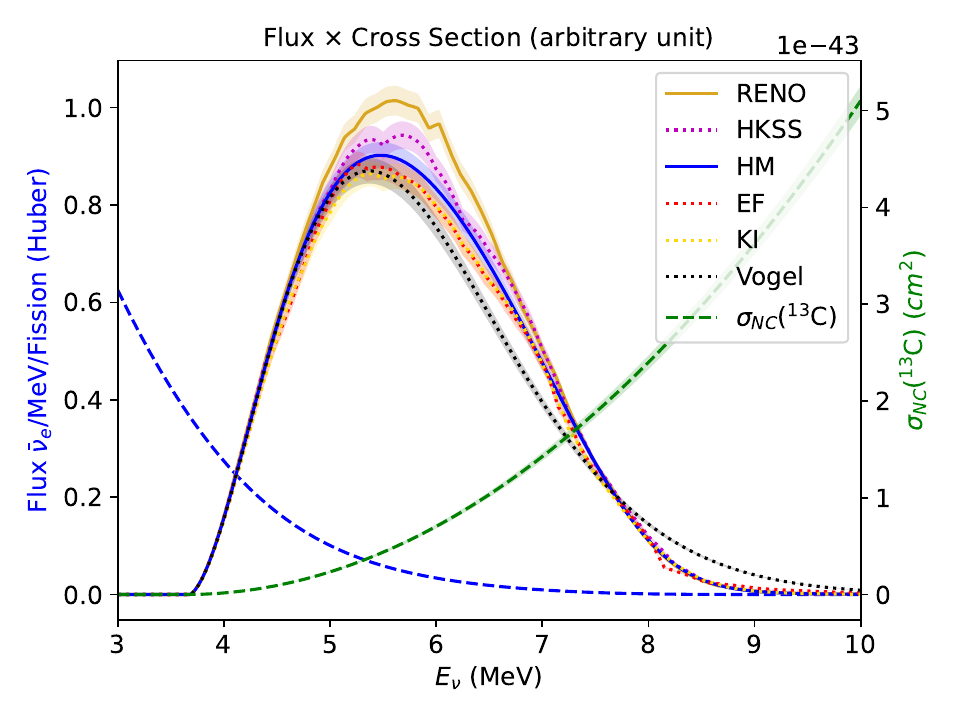}
\caption{
Blue solid: the expected number of events of our inelastic $\bar \nu_e - {}^{13}$C process obtained from the flux of HM model (blue dashed) and the cross-section (green dashed).
Yellow solid: assuming the flux is what is observed in RENO.
Others: assuming other flux models.
Shaded bands indicate the $\pm 3\%$ normalization uncertainty of the ${}^{13}\mathrm{C}$ neutral–current inelastic cross-section, which we expect to be achieved as discussed in Sec.~\ref{sec:xs}.
Because the event spectra scale linearly with $\sigma_{NC}$, the same relative $\pm 3\%$ band applies to all $\sigma\times$flux curves; no band is drawn for the flux-only (blue dashed) reference.
}
\label{fig:events}
\end{centering}
\end{figure}

The cross-section for the NC interaction of neutrinos
with $^{13}$C is given by~\cite{Suzuki:2012aa}:
\begin{equation}
\label{eq:C13cs}
\sigma(E_{\nu}) = \left[a_1 (E_\nu - Q) + a_2 (E_\nu - Q)^2 \right] \times 10^{-44}  \text{cm}^2\,,
\end{equation}
where $Q = 3.685$ MeV is the energy threshold, $a_1 = 0.122$, and $a_2 = 1.26$ are constants~\cite{Fukugita:1988hg,Suzuki:2019cra}.
Figure~\ref{fig:events} shows the cross-section represented by the dashed green curve.

The flux of reactor neutrinos is displayed with the dashed blue curve, assuming the HM model~\cite{Huber:2011wv, Mueller:2011nm} and considering the average fission fractions at near detector of RENO~\cite{RENO:2020dxd} as follows: 0.571 for $^{235}$U, 0.073 for $^{238}$U, 0.3 for $^{239}$Pu, and 0.056 for $^{241}$Pu. 
Other reactor models are not illustrated as they overlap with each other. 
The blue solid line is the flux (blue dashed) $\times\bar{\nu}_e$ interaction cross-section with $^{13}$C (green dashed) which is proportional to the expected number of events of our channel.
We also show the flux $\times$ the cross-section considering other flux models such as Vogel \cite{Vogel:1989iv}, KI (Kurchatov Institute) \cite{Kopeikin:2021ugh}, EF (Estienne-Fallot) \cite{Estienne:2019ujo}, and HKSS (Hayen-Kostensalo-Severijns-Suhonen) \cite{Hayen:2019eop}, as well as the reconstructed number of events considering the observed excess
at RENO~\cite{RENO:2015ksa}, for a review of different models, see \cite{Giunti:2021kab, Zhang:2023zif}.

As demonstrated in Fig.~\ref{fig:events}, a significant number of events falls within the neutrino energy range of 4.5 to 7.5 MeV, coinciding with the 5 MeV bump. This interaction thus presents a novel opportunity to investigate 
the origin of the bump, 
offering a new channel testing the phenomenon. 
The events ratios for the Vogel, KI, EF, and HKSS models to the HM model are 0.96, 0.96, 0.97, and 1.03, respectively, note that the HKSS model is introduced to fit the 5 MeV bump.
Assuming all the 
excesses
in the IBD measurements as reported in Ref.~\cite{Zhang:2023zif} are caused by a higher flux, 
the experimental observations of our alternative channel should also be
8$\%$ higher than 
the HM prediction.
Hence the reduction of the $\bar \nu_e {}^{13}{\rm C} \to \bar \nu_e {}^{13}{\rm C}^\ast$ cross-section uncertainties would help thoroughly investigating the high energy reactor neutrino flux, although we actually find the current level of uncertainties is still tolerable in probing certain explanations of the bump as discussed in the next section.
Before discussing further details, we just show the 3\% normalization uncertainties which are expected to be achieved by our proposals later in Fig.~\ref{fig:events} with shaded bands.

\begin{table}
\centering
\begin{tabular*}{\linewidth}{@{\extracolsep{\fill}}cccccc}
\hline
Experiments & P(GW) & m(t) & Baseline(m) &   events/yr\\ \hline \hline
Daya Bay (near/far)  & 17.4/17.4 & 80/80 & 578.7/1638  & 88/11 \\
RENO (near/far)  & 16.4/16.4 & 16/16 & 420/1400   & 38/3.4 \\
PROSPECT-II  & 3 & 4.8 & 25   & 491 \\ 
JUNO-TAO  & 4.6  & 2.8 & 30  & 305 \\
NEOS  & 2.8 & 1 & 24    & 103 \\ \hline
\end{tabular*}
\caption{Power, mass, baseline and the expected events. 
}
\label{table:Current Experiments}
\end{table}

We present the annual number of $^{13}$C events 
in
the reference 
reactor neutrino experiments in Table~\ref{table:Current Experiments}. The number of events is directly proportional to reactor power, detector mass, and the inverse of the square of the baseline. Assuming the average fission fraction reported by the near detector of RENO \cite{RENO:2020dxd} and employing the HM model, the annual number of $^{13}$C events in a commercial reactor is estimated to be approximately $22 \times (\text{Power/GW}) \times ({\rm Mass / kton}) / \left(\text{Baseline/km}\right)^2$.

\section{Test of New Physics}
\label{sec:NP}

\begin{figure}[tbp]
\begin{centering}
\includegraphics[width=\linewidth]{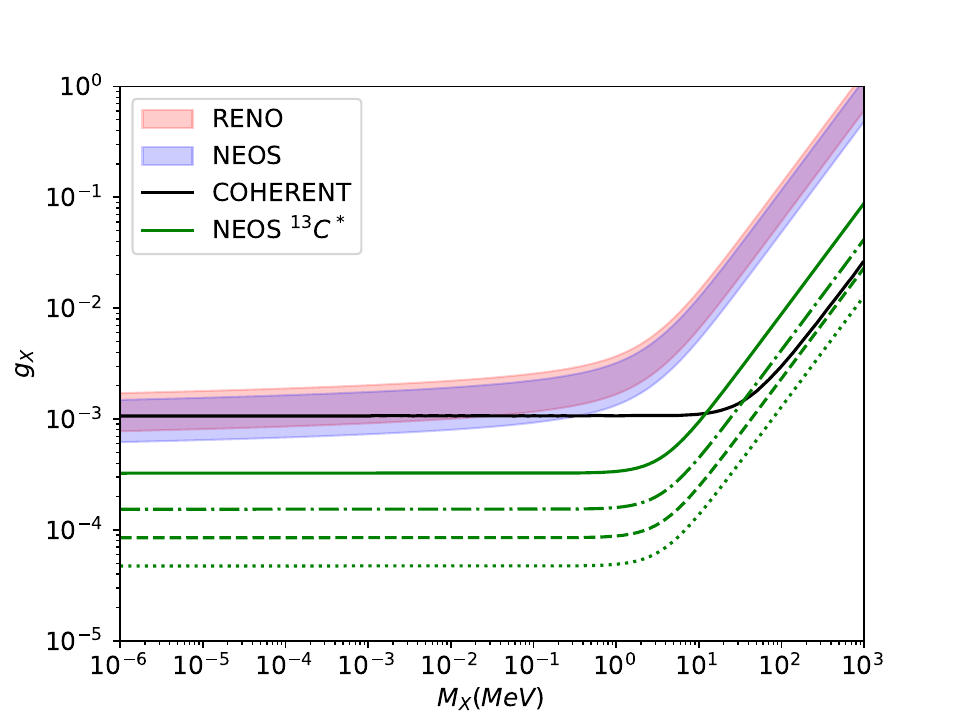}
\caption{
The $99\%$ C.L. parameter region explaining the 5 MeV bump observed in RENO and NEOS by 
$\nu_s {}^{13}{\rm C} \to \nu_s {}^{12}{\rm C}^\ast + n$
mediated by a new gauge boson $X$ with mass $M_X$ and the coupling $g_X$ with $\nu_s$ is completely ruled out by investigating possible sharp photon peak around 3.685 MeV from the 180 days of NEOS data (above the green solid).
}
\label{fig:sterile}
\end{centering}
\end{figure}

Interestingly, our alternative channel can efficiently test the new physics scenario explaining the 5 MeV bump~\cite{Berryman:2018jxt} {\it even without extra background subtraction} from the existing data of the NEOS experiment~\cite{Siyeon:2017tsg}.
The 5 MeV bump could arise from new physics involving either flux modifications (e.g., new particles decaying into \(\bar{\nu}_e\)) or mechanisms that do not alter the flux. 
Our $^{13}$C signal is able to test the latter case.
As a reference model, we consider the only BSM scenario proposed so far to explain the bump~\cite{Berryman:2018jxt} and use the NEOS data to conclusively rule out the vector interaction scenario.
The scenario considers a sterile neutrino $\nu_s$ with an additional U(1)$_X$ gauge symmetry, where the $\nu_s$ scattering with \(^{13}\text{C}\) 
mediated by the new gauge boson $X$
produces \(^{12}\text{C}^*\) and a neutron, mimicking the IBD signal. 
Intriguingly, the same interaction can also produce the \(^{13}\text{C}^*\) de-excitation signal, which leaves a sharp 3.685 MeV photon peak.
Our result is in Fig.~\ref{fig:sterile} where the preferred parameter region explaining the observed 5 MeV bump in RENO and NEOS is completely ruled out, i.e, it is above the upper limit from the 180 days of NEOS data~\cite{Siyeon:2017tsg} (green solid). In Ref.~\cite{Berryman:2018jxt}, the authors show possible constraints from the neutrino-induced deuterium disintegration performed in 1980. Note that our method can provide complementary constraint if the deuterium scattering experiment is re-performed in the near-future.

This is due to the fact that the inelastic $\bar \nu_s$ - ${}^{13}$C scattering mediated by the light gauge boson $X$ 
with at most sub-GeV mass is at least million times larger than the SM process via $Z$ boson exchange.
Extra data collection for 10 years (green dot-dashed) as well as more dedicated background reduction to the level comparable to the number of signals (dashed for 180 days and dotted for 10 years), following the strategies explained in the next section, 
increases the sensitivity. 
More detailed explanation of how we obtained the expected signals from the vector mediator scenario is provided in Appendix~\ref{appendix:nus}. Technically, Ref.~\cite{Berryman:2018jxt} also includes an axial vector mediator and magnetic-magnetic interaction through the $Z'$ boson exchange. We naively expect that the axial vector mediator scenario inducing our 3.685 MeV photon channel can be also excluded similarly due to the huge difference in the cross-section compared to the SM process, although not concrete analysis has been made here. For the magnetic dipole moment interaction, dedicated study is needed.
Note that our method is generically applicable to new physics scenarios that do not alter the flux itself,
which is a critical point to emphasize.

\section{Cross-section measurement}
\label{sec:xs}

Reduction of the systematic uncertainties can extremely increase the power of our alternative channel to identify the origin of the bump beyond the specific example or similar level BSM contributions discussed in the previous section. 
If the uncertainty is below 8\%, we can even separate some of the reactor neutrino flux models.
A major source of systematic uncertainties is the cross-section.
For the first time, we investigate 
the potential of future experiments in precisely determining the $^{13}$C cross-section. We propose a novel approach using the $^8$B solar neutrinos at JUNO, pion and muon decay at rest ($\pi$ DAR and $\mu$ DAR), and isotope decay at rest (IsoDAR). 

The cross-section detailed in Eq.~(\ref{eq:C13cs}) obtained from a shell model Hamiltonian has not yet been verified from experiments.
Although about 10\% uncertainty is conventionally expected in the nuclear physics calculation so far, further developments such as the ab initio nuclear structure calculations incorporating chiral three-nucleon interactions and considering results from LSND, Super Kamiokande, and other neutrino experiments~\cite{Barrett:2013nh}, might lead to more precise calculations. 
The JUNO collaboration plans to measure the $^{8}$B solar neutrino flux via $\nu {}^{13}{\rm C} \to \nu {}^{13}{\rm C}^\ast$, with the hope of the precise cross-section calculation below 1\% in the future~\cite{JUNO:2022jkf}, which may offer promising opportunities for measuring the \(^{8}\text{B}\) solar neutrino flux.

In contrast, we pursue to utilize the current and future experiments to reduce the cross-section uncertainties in this paper, without adopting the potential of precise cross-section calculation which is not proceeded yet.
As studied in Refs.~\cite{Arafune:1988hx, Ianni:2005ki, JUNO:2022jkf}, the ${}^8$B solar neutrino measurements via the $^{13}$C NC interaction are possible in the LS detectors.
The current $^8$B flux uncertainty is 3.5\% (SNO)~\cite{SNO:2003bmh}. 
However, the subpercent level measurements of \(\Delta m^2_{21}\) and \(\theta_{12}\) by reactor neutrinos in JUNO and $\nu$EYE in Yemilab, along with the detection of pp, $^7$Be, and pep solar neutrinos there~\cite{Bakhti:2023vzn}, can precisely determine the $^8$B flux below 1\% precision, in combination with the ${}^8$B solar neutrino measurements in Hyper-Kamiokande (HK)~\cite{Hyper-Kamiokande:2018ofw}, DUNE~\cite{Capozzi:2018dat,Bakhti:2020tcj}, and THEIA~\cite{Theia:2019non} via ES and CC neutrino interactions. 
Also, the dark matter direct detection experiments are expected to achieve 1 \% level precision of the flux from observing the CE$\nu$NS events~\cite{Cerdeno:2016sfi}. 
Based on these expectations, we analyze the prospect of JUNO in measuring $\sigma(\nu {}^{13}{\rm C} \to \nu {}^{13}{\rm C}^\ast)$ assuming the flux of the $^8$ B solar neutrino within 1\% precision.

Neutrino production from $\pi$ DAR and $\mu$ DAR in experiments like LSND~\cite{LSND:2001aii}, JSNS$^2$~\cite{JSNS2:2021hyk}, and OscSNS which is a proposal to search for sterile neutrinos in Oak Ridge National Lab.~\cite{OscSNS:2013vwh} with precise information of the neutrino flux about 0.8\% can be used to measure the cross-section at a few tens of MeV energy range.
Based on the data-taking volume similar to that of LSND, we expect approximately 13 events of $^{13}$C excitation from neutrinos produced via pion decay at rest (with a monochromatic energy of $30$ MeV), while 40 events from neutrinos can be produced via muon decay at rest. 
JSNS$^2$ is expected to collect a number of events comparable to those observed by LSND, both of which have a potential to measure the cross-section with 15\% precision assuming 10\% of the total systematic uncertainty.
On the other hand, a future proposal of the OscSNS (or similar level of next generation $\pi/\mu$ DAR experiments) using 886 tons of mineral oil and a total neutrino flux of $1.64 \times 10^{14} \nu/(\text{year} \cdot \text{cm}^2)$, and assuming three years of data collection, we anticipate collecting events at a rate twenty times higher than what was observed in the LSND experiment.

Considering $\pi$ DAR and $\mu$ DAR with neutrino energies beyond 10 MeV, the cross-section becomes sensitive
predominantly to the $a_2$ term of Eq.~(\ref{eq:C13cs}). The $a_1$ term is negligible, accounting for less than $0.3\%$ of the total cross-section. Assuming theoretical uncertainty of $10\%$ for both $a_1$ and $a_2$, the effect of $a_1$ on the total cross-section uncertainty is approximately $0.2\%$ in the case of reactor neutrinos with energy range below $\lesssim 10$ MeV.
Furthermore, a combination of lower-energy neutrinos from solar and isotope decay at rest sources with higher-energy neutrinos from $\pi$DAR and $\mu$DAR could help constrain the cubic term $a_3$, which is theoretically set to zero, as well as higher-order terms. 
The analysis including the subdominant $a_3$ term is discussed at the end of this section.

The $\nu$EYE at Yemilab can detect solar, reactor and IsoDAR-produced neutrinos from a high-purity proton beam with energy 60 MeV.
The IsoDAR, with $1.97 \times 10^{24}$ protons on target, will produce $\bar \nu$ from $^8$Li beta decay ($Q = 16$ MeV)~\cite{Alonso:2021kyu,Seo:2023xku}.
The details including the fiducial volume and the exposure time we assumed to obtain our ${}^{13}$C events in the reference experiments are summarized in Table~\ref{table:futurecross}. 
Note that JUNO-TAO assuming the flux measured in RENO via IBD is included to show its capability.

\begin{table}[h]
\centering
\begin{tabular*}{0.5\textwidth}{@{\extracolsep{\fill}}ccccc}
\hline
Experiments & Fiducial vol. (t) & Exposure time (yr) &   events\\ \hline \hline
JUNO  & 12.2k & 10  & 1800 \\
JUNO-TAO  & 2.8 & 10  & 3000 \\
OscSNS  & 886 & 3  & 1000 \\
IsoDAR  & 1.2k & 4 & 320 \\ \hline
\end{tabular*}
\caption{Fiducial volume, exposure time, and the expected number of $\nu$ - $^{13}$C NC events in the  reference experiments.}
\label{table:futurecross}
\end{table}

Sensitivity studies from solar neutrino experiments like JUNO, reactor experiments like JUNO-TAO, over ten years of data-taking, OscSNS $\pi$/$\mu$ DAR experiment and IsoDAR at Yemilab, are illustrated in Fig.~\ref{fig:sensitivity_a2}.  
In obtaining the sensitivities, we optimistically estimate the signal-to-background ratios of \(15/1\) for JUNO-TAO, \(2/1\) for JUNO, and \(1/1\) for IsoDAR at Yemilab, while the ratio \(1/1\) is assumed for OscSNS without performing dedicated analysis, as will be explained later. 

\begin{widetext}

For JUNO and JUNO-TAO, neglecting the background, collecting more than 1800 and 3000 events, each experiment can achieve better than $2.5\%$ and $2\%$ sensitivity in measuring the cross-section independently. For the case of OscSNS, collecting one thousand of events, measure the cross-section better than $3\%$.  Taking the background into account, the precision of JUNO and OscSNS is $3\%$ and $4\%$, respectively. Notably, these projections only account for statistical uncertainties in the cross-section measurement.
\begin{figure}[tbp]
\begin{centering}
\includegraphics[width=8.0cm]{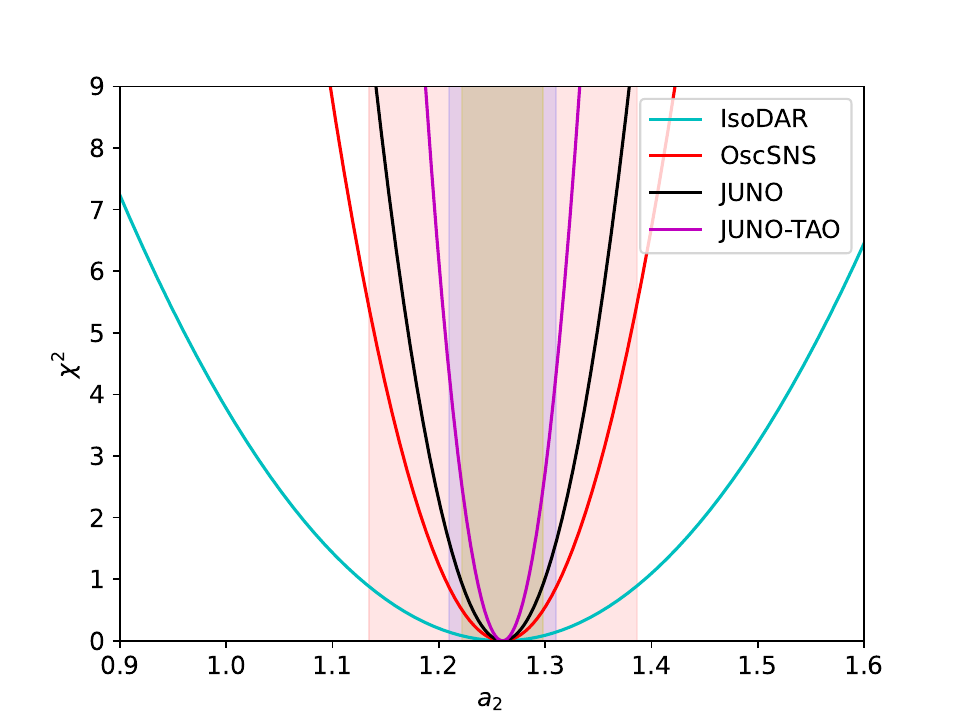}
\includegraphics[width=8.4cm]{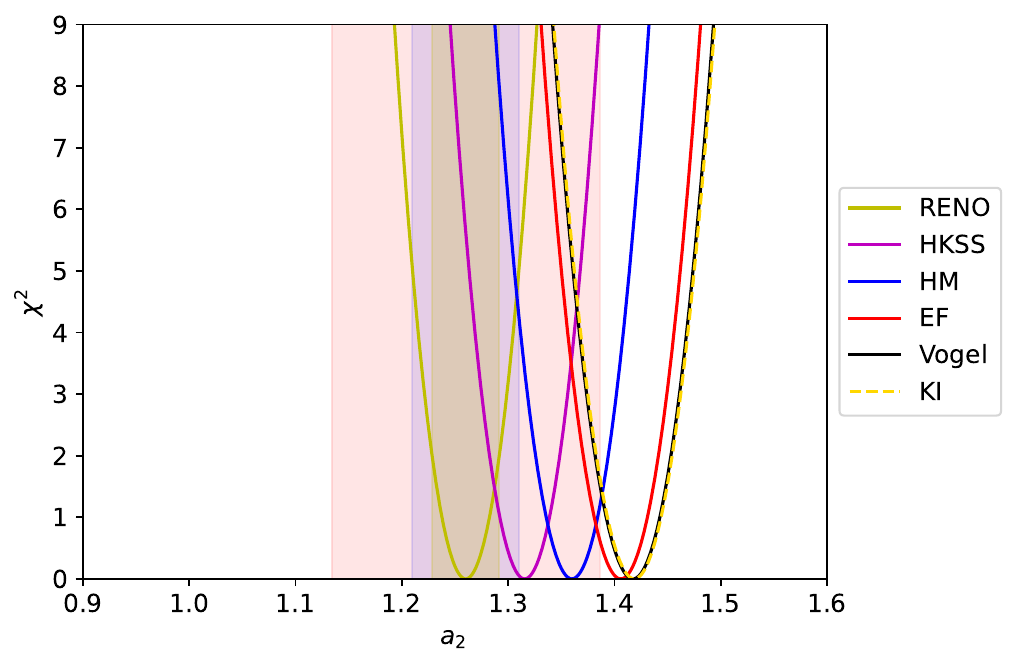}
\caption{ $\chi^2$ as a function of the quadratic term constant $a_2$ in Eq.~(\ref{eq:C13cs}), the left panel shows the result for different experiments. The right panel shows the same for different reactor models, using JUNO-TAO experiment. 3\%, 4\%, and 10\% precision measurements indicated by the shaded grey, blue, and pink regions, respectively, corresponds to JUNO, OscSNS and IsoDAR precision at 1$\sigma$.
Assuming semi-experimental flux from the excess at RENO as the true model and $a_2 = 1.26$, the measured values of $a_2$ are 1.31, 1.36, 1.41, 1.42, and 1.42 for the HKSS, HM, EF, Vogel, and KI models, respectively.
The difference between the reconstructed RENO flux and theoretical models ranges from 4\% to 12\%.
}
\label{fig:sensitivity_a2}
\end{centering}
\end{figure}
\end{widetext}

In the process of background estimation, we impose the anti-coincidence cut to reject the IBD events with an efficiency of 99.9\%, following the study in Ref.~\cite{Conrad:2004gw}.
The background events coming from the $\nu-e$ elastic scattering ($\nu$ES), radioactivity, and isotopes are scaled from Refs.~\cite{Conrad:2004gw, JUNO:2022jkf, Alonso:2021kyu}.
Following the expectations in JUNO and JUNO-TAO, we assume a low level of ${}^{232}$Th contamination rate at most $5 \times 10^{-17}$g/g, providing  around one \(^{208}\text{Tl}\) beta decay background event within the region of interest (ROI) per day in a 1 kt LS detector. 
One can further reduce the 99\% (JUNO) and 80\% (JUNO-TAO) radioactive backgrounds, respectively, by ${}^{232}$Th chain tagging~\cite{JUNO:2020hqc, Hachiya:2020mpu}, making them negligible compared to the $\nu$ES background events, while we do not consider such a possibility for conservativeness here.
Notably, we adopt the 95\% level $\beta/\gamma$ separation efficiency in our ROI. 
This is quite challenging for most of the LS detectors but recently some JUNO collaboration members studied that 90\% level $\beta/\gamma$ separation efficiency in the energy range of 1.25$-$1.75~MeV is achievable with the help of the neural network and the topological reconstruction in JUNO and JUNO-TAO~\cite{Rebber:2020xfi}, also, a future detector design LiquidO is expected to achieve even better efficiency~\cite{LiquidO:2019mxd}, although this is not a LS detector.
Conventionally the discrimination efficiency increases with energy and hence we expect the 95\% level $\beta/\gamma$ separation efficiency would be feasible. Note that the use of slow LS, incorporating both Cherenkov and scintillation light, makes it available to determine the direction of the background events such as $\nu$ES electrons~\cite{Biller:2020uoi}, and hence improve our sensitivities by further background rejection. 
The details of our expected signals, selection cuts, and the remaining background events in JUNO-TAO, JUNO, and $\nu$EYE-IsoDAR@Yemilab are listed in Table~\ref{table:eventsandbackgrounds}.

Note that JUNO with conventional fiducial volume of 20 kt has a capability of observing 300 signal events per year approximately. However, due to additional cuts implemented to reduce environmental backgrounds, the effective fiducial volume is reduced to 12.2 kt in our case~\cite{JUNO:2020hqc}. The main sources of background are $\nu$-e scattering, environmental background from radioactive isotopes, reactor events, and $\nu_e$CC interactions. Assuming a 95\% $\beta/\gamma$ discrimination and a 3\% energy resolution, we expect that $\nu$-e scattering is the most significant background source before applying the $\beta/\gamma$ separation cut, contributing approximately one-third of the total background events after all the cuts. 
\begin{widetext}
Overall, we estimate the number of background events after the cuts in JUNO to be half of the signal, leading to 3\% precision as in Fig.~\ref{fig:sensitivity_a2}.

\begin{table}
\begin{minipage}{\textwidth}
\centering
\begin{tabular*}{\textwidth}{@{\extracolsep{\fill}}c|cccc}
\hline
Experiment & JUNO-TAO & JUNO & $\nu$EYE-IsoDAR at Yemilab\\ \hline \hline
Fiducial volume  & 2.8 t & 12.2 kt & 1.2 kt  \\
Reactor Power (GW) & 4.6 & 26.6  & 25 \\
Reactor Baseline  & 28 m & 53~km & 65~km \\ 
Accelerator Baseline (m) & N/A & N/A  & 17 \\ 
Proton-On-Target & N/A & N/A & $1.97\times10^{24}$ \\
ROI & 3.685$\pm0.02$ MeV & 3.685$\pm0.1$ MeV & 3.685$\pm0.1$ MeV \\ 
Assumed $^{232}$Th contamination & $5\times10^{-17}$g/g & $10^{-17}$g/g & $5.7\times10^{-19}$g/g \\ \hline
reactor $\bar{\nu}-{}^{13}\text{C}$ NC & 300 & $\lesssim$ 1  & $\lesssim1$  \\ 
solar $\nu-{}^{13}\text{C}$ NC & $\lesssim 1$ & 180 & 18 \\
IsoDAR $\bar{\nu}-{}^{13}\text{C}$ NC & N/A & N/A & 80 \\ \hline
$\nu(\bar{\nu})$ES & 330 & 540 & 370 \\
IBD & 50 & $< 1$ & 20 \\
radioactivity & $\lesssim 1$  & 60 & 60 \\ \hline
Background after 95\% $\beta/\gamma$ separation & 20 & 87 & 80 \\ \hline
Total number of signal & 300 & 180 &  80\\ \hline
\end{tabular*}
\caption{
Annual number of events and backgrounds in the ROI for our reference reactor, solar, and IsoDAR neutrino sources at organic liquid scintillator detectors are shown. 
Note again that JUNO-TAO, assuming the reactor neutrino flux measured in RENO via IBD is exact, is included just to show its capability in measuring the $\bar \nu_e - {}^{13}$C cross-section.
Power and Baseline in the table refer to the effective nuclear reactor power and baseline for each detector. Following~\cite{Conrad:2004gw}, 99.9\% of IBD backgrounds are assumed to be rejected after anti-coincidence cut. $\nu$ES, radioactivity and isotope backgrounds are scaled from~\cite{Conrad:2004gw, JUNO:2022jkf, Alonso:2021kyu}. Note that assumed $^{232}$Th level for the JUNO and JUNO-TAO is solar neutrino detector level radiopurity, and we further note that they can reduce 99\% and 80\% of radioactive background respectively by $^{232}$Th tagging~\cite{JUNO:2020hqc, Hachiya:2020mpu}. For $\nu$EYE-IsoDAR scenario, we assumed that the energy resolution is good enough that most of the signals reside in the ROI. Note that we assumed additional fiducial volume cut for JUNO and $\nu$EYE to reduce radioactive background from the detector wall and outside of the detector following~\cite{JUNO:2020hqc,Alonso:2021kyu}.}
\label{table:eventsandbackgrounds}
\end{minipage}
\end{table}
\end{widetext}

For $\nu$EYE - IsoDAR at Yemilab with 2.26 kt fiducial volume originally, we implemented an additional volume cut to mitigate radioactive background from the detector wall and external sources, as described in Ref.~\cite{Alonso:2021kyu}.
The number of background events can be reduced to that comparable to the signal events if 95\% level $\beta / \gamma$ separation is achievable in the IsoDAR at Yemilab~
\cite{Alonso:2022mup}.
Then, we find the cross-section can be measured with the precision of 10\% as shown in Fig.~\ref{fig:sensitivity_a2}.
If the ${}^{232}$Th chain tagging is achievable and hence the signal-to-background ratio about one can be kept in the full fiducial volume of 2.26 kt, the $\bar \nu_e - {}^{13}$C cross-section precision becomes better than 7.5\%. If we can further reduce the background to negligible level compared to the signal events inside the full fiducial volume of 2.26 kt, the precision becomes 4\%.
Moreover, combined with the 34 events per year from $^8$B solar neutrinos in the same detector, the ten years of solar data collection and four years of beam-on data (IsoDAR), the $\nu$EYE at Yemilab can maximally obtain a statistical sensitivity of 
$3\%$ by collecting more than 900 events.
It is intriguing that the combination of both solar and IsoDAR data in JUNO and $\nu$EYE at Yemilab can lead to much better precision in the $\bar \nu_e - {}^{13}$C cross-section, which needs a dedicated study. Assuming perfect efficiency and negligible background compared to the signal events, we naively estimate the statistical sensitivity can be reduced to 1.5\% with the data collection of approximately 4000 signal events.

For OscSNS, we ambitiously assume the signal-to-background ratio of one, without detailed estimations.
The reason is due to the beam related or cosmic muon induced fast neutron backgrounds, which need dedicated investigation using the programs such as GEANT4~\cite{Allison:2016lfl} and rely much on the details of the detector designs.
On the other hand, the thermal neutron backgrounds can be mitigated with additional shielding easily.
Also, the environmental and accidental backgrounds can be significantly reduced considering the pulsed nature of the beam.

\begin{figure}[tbp]
\begin{centering}
\includegraphics[width=\linewidth]{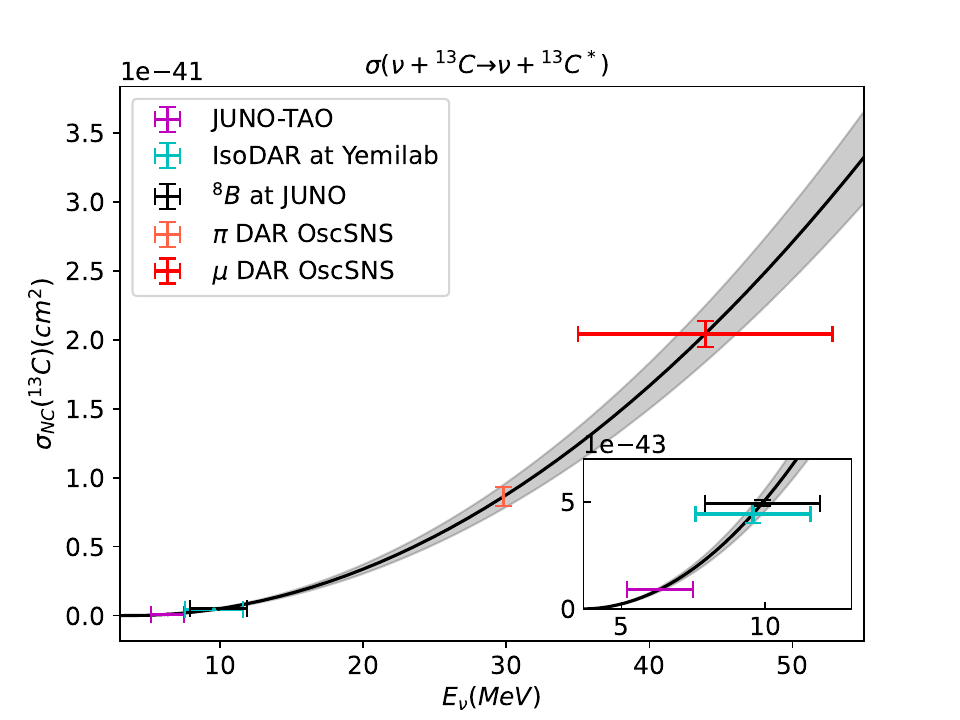}
\caption{The prospects of measuring the cross-section in a wide range of neutrino energy in reference experiments.
}
\label{fig:sigmaE}
\end{centering}
\end{figure}

Our results of the cross-section measurements in a wide range of neutrino energy region from various future experiments are displayed in Fig.~\ref{fig:sigmaE}.
The theoretical cross-section in Eq.~(\ref{eq:C13cs}) is shown as the black solid line with 10\% uncertainty (gray shaded band) conservatively.
We conclude that JUNO (solar data) and $\nu$EYE - IsoDAR at Yemilab can measure the cross-section within 3\% and 10\% precision, respectively, in the energy region for $E_\nu \lesssim 10$ MeV, while OscSNS has sensitivity of 4\% in the high energy region. 
Note that controlling the systematic uncertainties in those experiments is very important to achieve our estimation.
It is worth to emphasize that further background mitigation would increase the sensitivities, e.g., up to 4\% for IsoDAR@Yemilab with 2.26 kt fiducial volume and negligible background, requiring dedicated efforts.

\begin{figure}[tbp]
\begin{centering}
\includegraphics[width=8.0cm]{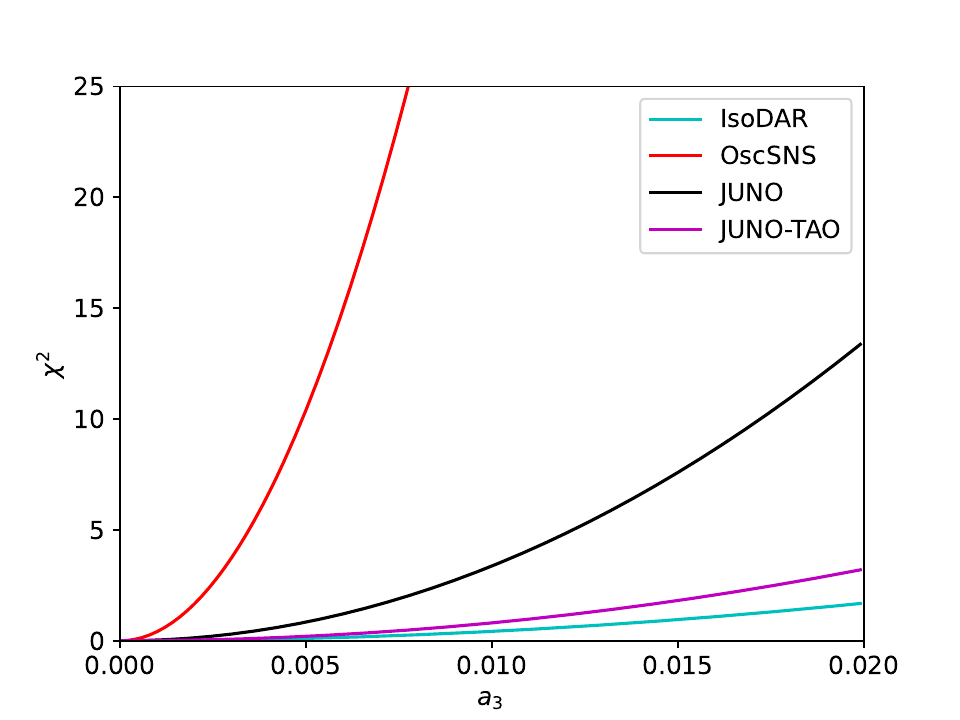}
\includegraphics[width=8.4cm]{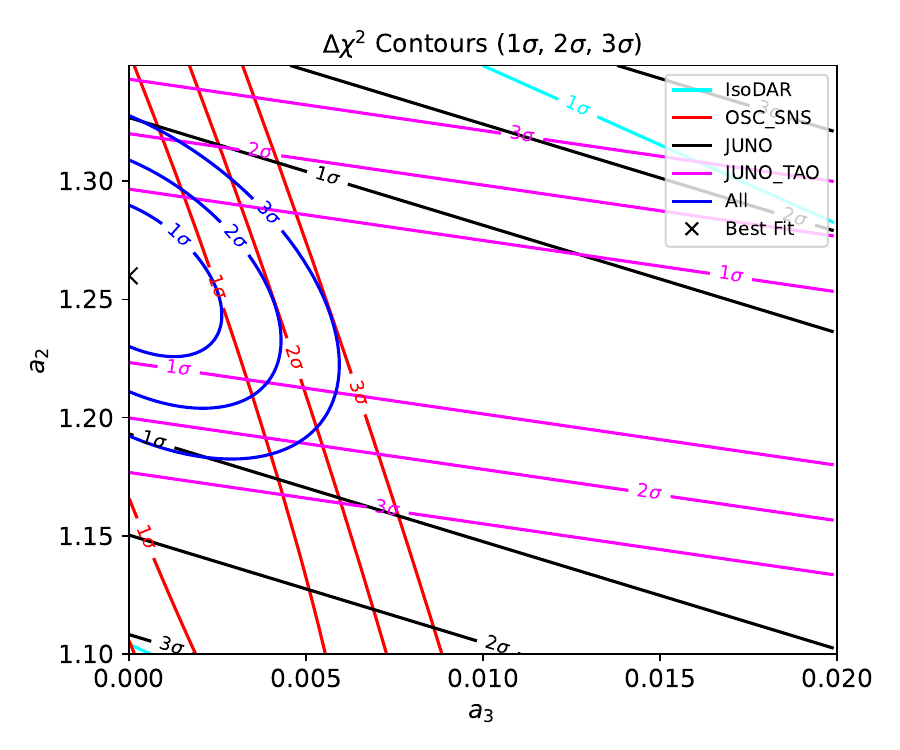}
\caption{\textbf{Left:} $a_{3}$ versus $\chi^{2}$ profiles for IsoDAR, OscSNS, JUNO, and JUNO-TAO, obtained with fixed $a_{2}$ at each experiment’s best-fit value. The corresponding $1\sigma$ upper bound on $a_{3}$ are: IsoDAR $=0.015$, OscSNS $=0.0016$, JUNO $=0.0055$, JUNO--TAO $=0.0066$. 
\textbf{Right:} Joint $(a_{2},a_{3})$ confidence regions for each experiment and for the combined fit; the $1\sigma,\,2\sigma,\,3\sigma$ contours for two degrees of freedom. Stars indicate best-fit points. For reference, the one-parameter $\chi^{2}(a_{2})$ profiles are shown in Fig.~\ref{fig:sensitivity_a2}.}
\label{fig:a3_chi2_contours}
\end{centering}
\end{figure}

Before closing the session, we further analyze the effect of adding the subdominant cubic term in the cross-section in Eq.~(\ref{eq:C13cs}), whose central value is set to zero in Ref.~\cite{Suzuki:2012aa}. 
We then perform both single-parameter and simultaneous fits to four complementary data sets: solar neutrinos at JUNO, reactor $\bar{\nu}_{e}$ at JUNO–TAO, and decay-at-rest (DAR) neutrinos at IsoDAR and OscSNS. The upper panel of Fig.~\ref{fig:a3_chi2_contours} shows the one-parameter sensitivities to $a_{3}$ with $a_{2}$ fixed at its best-fit value, while the lower panel displays the joint $(a_{2},a_{3})$ confidence regions (1–3$\sigma$) for each experiment and for their combination. 
Individually, the experiments constrain different linear combinations of $(a_{2},a_{3})$ due to their distinct neutrino energy spectra; the combined fit breaks these degeneracies and yields the most stringent sensitivities across the full energy range.

The resulting one-parameter $1\sigma$ sensitivities on $a_{3}$ are: IsoDAR $0.015$, OscSNS $0.0016$, JUNO $0.0055$, and JUNO–TAO $0.0066$. Among these, OscSNS provides the best sensitivity because of its higher neutrino energies, where the cubic term has a greater impact. 
The combination of experiments with different neutrino energy ranges enables a simultaneous determination of $a_{2}$ and $a_{3}$, since their sensitivities to the quadratic and cubic terms differ. 
The combined fit in the lower panel of Fig.~\ref{fig:a3_chi2_contours} yields $a_{2} = 1.26 \pm 0.03$ and an upper bound on $a_{3}$ of $0.003$ at $1\sigma$. 
For reactor neutrino experiments with $E_{\nu} < 10$ MeV (average $E_{\nu} \approx 5.8$ MeV), neglecting $a_{3}$ leads to only $\sim 0.5\%$ effect on the cross-section and total event rate at $1\sigma$.

\section{Flux measurement sensitivities}
\label{sec:fluxsens}

As mentioned in the introduction, the 5 MeV bump can be originated from unidentified IBD systematics in the neutrino energy region of 4--8 MeV. 
In addition, there can be BSM scenarios mimicking the IBD signal but not contributing to the 3.685 MeV photon signal, although not proposed yet.
In those cases, no distinctive bump over the theoretical expectation is expected in our $\bar \nu_e - {}^{13}$C channel in the 4--8 MeV region, unlike the observations via IBD. 
As displayed in the right panel of Fig.~\ref{fig:sensitivity_a2}, testing the aforementioned possibilities requires the capability of at least 12\% flux separation between the observed one in RENO (yellow solid) and theoretical models, e.g., KI model (yellow dotted) resolving the RAA.
In this section, we investigate the feasibility of obtaining such a flux separation sensitivity via our channel in various experiments by imposing the background subtraction strategies discussed in the previous section. 

For simplicity, we adopt the three scenarios based on the aforementioned background subtraction strategies: $N_{\rm background}/N_{\rm signal}$ = 6, 1, and 0.
The first scenario assumes that all the background events originate from the electron scattering of reactor neutrinos, with zero efficiency for $\beta/\gamma$ separation
but with low ${}^{232}$Th contamination at the level of $5 \times 10^{-17}$g/g and an overburden of $\mathcal O (100\,{\rm m})$. 
This assumption, while conservative for experiments like Daya Bay and RENO, is plausible for very short baseline experiments.
The second scenario is well-supported by 
additionally
considering effective $\beta/\gamma$ discrimination
around 90 - 95\%, along with significant reductions in muon and neutron backgrounds to match the signal level.
We expect JUNO and IsoDAR@Yemilab can have such capabilities of realizing this scenario, as discussed previously.
The third scenario is chosen to show the best-case scenario.
We expect very low level of background contamination compared to the signal events can be achievable in JUNO-TAO based on the studied in Refs.~\cite{Rebber:2020xfi,JUNO:2020ijm} and our results in the previous section.

\begin{figure}[tbp]
\begin{centering}
\includegraphicsifexists[width=\linewidth]{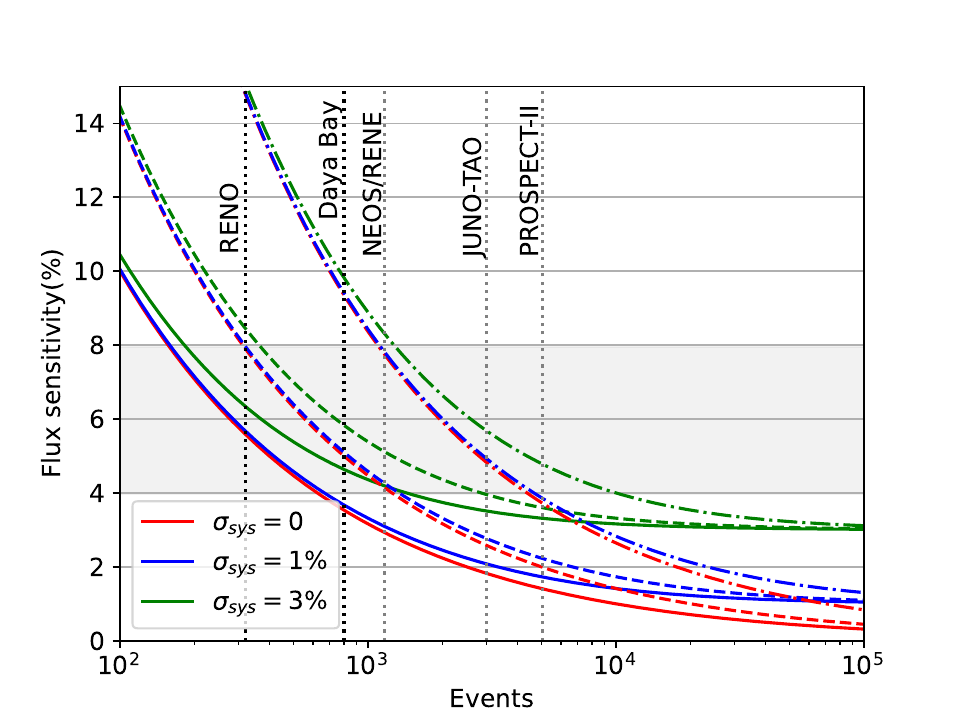}
\caption{
Signal acceptance is universally assumed to be 100\% for simplicity.
We put NEOS and RENE together although RENE can reduce $\gamma$ background with better $E$ resolution.
}
\label{fig:sensitivity}
\end{centering}
\end{figure}

In Fig.~\ref{fig:sensitivity}, the dotted, dashed, and solid curves represent the 1$\sigma$ sensitivities for 
the aforementioned
three background scenarios 
$N_{\rm bkg.}/N_{\rm sig.} = 6, 1, 0$, respectively, with red, blue, and green curves showing systematic uncertainties of 0\%, 1\%, and 3\%, respectively. The separation between these curves increases with smaller backgrounds. Vertical dotted lines indicate the statistics of past (RENO, Daya Bay), current, and future experiments, assuming 10 years of data for the latter. 
The light gray shaded region (4-8\%) is the level of flux measurement uncertainties that can separate the expected 5 MeV bump and the HM model and even different flux models as a byproduct.
Note that the observed bump can be confirmed compared to the EF and KI models when the uncertainty becomes below the 12\% level. 

It is remarkable that JUNO-TAO is expected to achieve the precision even below 4\% after the 10 years of running if its background subtraction capability based on the proposals~\cite{Rebber:2020xfi,JUNO:2020ijm} is kept. 
In addition, we expect the combined data of the NEOS and RENE (Reactor Experiment for Neutrino and Exotics)~\cite{Choi:2025wbw} can have sensitivities below 8\% , making the separation between the 5 MeV bump and the HM viable, if our extra neutron background subtraction strategy in Appendix~\ref{appendix:moresubtraction} is adopted. 

Note that the systematic uncertainty from the cross-section measurement can be at least 3\% (green lines) by nominally adopting our results in Sec.~\ref{sec:xs}.
However, additional ${}^{232}$Th chain tagging can increase the precision and hence the optimistic sensitivities can be somewhere between the green lines ($\sigma_{\rm sys} = 3$\%) and the blue lines ($\sigma_{\rm sys} = 1$\%).
If additional systematic uncertainties exist or the cross-section precisions of 3\% is not experimentally obtained, along with a partial failure of the background subtraction as explained in Ref.~\cite{JUNO:2020ijm} and our previous section, then the sensitivity of JUNO-TAO after the 10 years of running should be between the green solid line (signal-to-background $\ll$ 1 and $\sigma_{\rm sys} = 3$\%) and the green dot-dashed line (signal-to-background = 6, and $\sigma_{\rm sys} = 3$\%), where we still expect the sensitivities of 5 to 6\%.
Even then, we expect JUNO-TAO would have capabilities of separating the 5 MeV bump and the HM model.

\begin{figure}[tbp]
\begin{centering}
\includegraphics[width=\linewidth]
{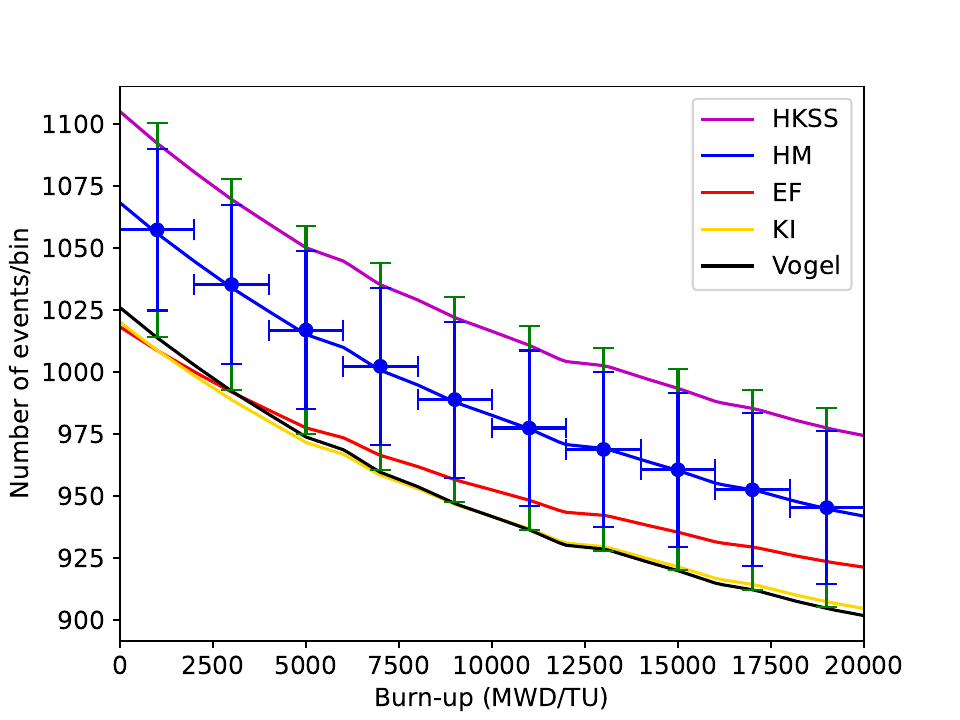}
\caption{
Number of events expected from each flux model versus burn up in megawatt-day over thermal unit~(MWD/TU) at an experiment collection ten thousand events, considering Asimov dataset and HM as the true model.  }
\label{fig:evolution}
\end{centering}
\end{figure}

Now we discuss another powerful method investigating the 5 MeV bump, the fission fraction evolution, which is less affected by the systematic uncertainties.
As can be observed from Fig.~\ref{fig:evolution}, assuming fission fractions evolution in Daya Bay~\cite{DayaBay:2016ssb}, we can even distinguish some flux models by observing the time evolution of the reactor. The blue and green error bars correspond to 
$1\sigma$
statistical only and statistical plus 1\% systematic uncertainties respectively, assuming collection of ten thousand events. 
Without considering systematics and background, it is expected to achieve a 3.5$\sigma$ sensitivity in discriminating between the HKSS and HM models. Considering 1\% and 3\% systematic uncertainties and including background comparable with signal, such an experiment can distinguish between HKSS and KI at 4.5$\sigma$ and 3.5$\sigma$ respectively, while it discriminates between HKSS and HM at 2.5$\sigma$ to 1.5$\sigma$. 
The near-future experiments JUNO-TAO and PROSPECT-II have 3$\sigma$ and 4$\sigma$ sensitivities, respectively, to distinguish between HKSS and KI models assuming 1\% systematics and neglecting the background.

Another crucial question is how much different isotopes contribute to this bump~\cite{Huber:2016xis,DB:2019U235U239,RENO:2019Fuel,DBPROSPECT:2022Joint,DB:2023FuelEvol,Zhang:2024PPNP}. 
The IBD measurements across multiple baselines and fuel compositions (Daya Bay, RENO, Double Chooz, NEOS, PROSPECT, STEREO) have shown that the 5 MeV bump appears in association with several fission isotopes rather than being attributable to a single source, and fuel-evolution analyses provide quantitative constraints on the relative roles of $^{235}$U and $^{239}$Pu~\cite{Huber:2016xis,RENO:2019Fuel,DB:2019U235U239,DBPROSPECT:2022Joint,DB:2023FuelEvol,Zhang:2024PPNP}.
Since the $^{239}$Pu fission fraction increases in time, by observing $^{13}$C events during fuel evolution, one can find the contribution of different isotopes to the 5~MeV bump~\cite{RENO:2019Fuel,DB:2023FuelEvol}.  
Compared to the IBD events, a steeper (shallower) decrease in the number of events indicates a larger contribution from $^{239}$Pu ($^{235}$U) to the bump~\cite{RENO:2019Fuel}.
A combination of IBD and $^{13}$C will make it more accessible to investigate the contribution of the different isotopes to the 5 MeV bump.

This indicates the potential to observe fuel evolution within a reactor, as well as to measure fission fractions during the fuel cycle or the beta spectrum ratio between $^{235}$U and $^{239}$Pu~\cite{DB:2019U235U239,DBPROSPECT:2022Joint}. 
Notice that the fission fraction of $^{238}$U remains nearly constant, while that of $^{241}$Pu exhibits a roughly linear relationship with the $^{239}$Pu fraction~\cite{DB:2023FuelEvol}.

\section{Conclusions}
\label{sec:conclusions}

Since its first observation in 2015, the 5 MeV bump has been confirmed in all the reactor experiments so far.
Notably, the advanced flux models which can resolve the RAA such as KI and EF models, in contrast, exacerbate the 5 MeV bump.
The bump may originate from a miscalculation of the flux, IBD systematics, or BSM contributions.
In light of this situation, we emphasize that suggesting a new reactor neutrino observation channel other than IBD would be extremely helpful to identify the origin of the 5 MeV bump and clarify the high energy ($\gtrsim$ 4 MeV) reactor neutrino flux. 
As such an alternative channel, we proposed the inelastic scattering of neutrino with $^{13}$C emitting a distinctive 3.685 MeV photon in the process of $^{13}$C$^\ast$ de-excitation after the scattering.
The comparison of the observations between the IBD and our channel can help identify the origin: 
i) a miscalculation or modification (by BSM) of the flux if the bumps are comparable, 
ii) IBD systematics or BSM contributions that mimic only the IBD (but not our 3.685 MeV photon channel) if no-bump is observed in our channel, and 
iii) a BSM contribution, possibly containing light new particles, both mimicking the IBD and inducing a 3.685 MeV photon peak if the 5 MeV bump {\it via our channel} becomes more pronounced.

The strength of our proposal is confirmed by our analysis of the existing NEOS data (collected via IBD) which exclude at $>99\%$ confidence level, within the benchmark parameter space considered, the BSM scenario used to explain the 5 MeV bump: a sterile-neutrino framework with a light vector mediator $Z'$ that couples to quarks and neutrinos.  
We nominally applied the (null) result in the energy window of 3.685 $\pm$ 0.1 MeV in the announced NEOS data without additional background subtraction.
This belongs to our third category mentioned above, i.e., a BSM contribution both mimicking the IBD and inducing a 3.685 MeV photon peak.
Moreover, any BSM explanations of the 5 MeV bump that include light-mediator interactions with ${}^{13}$C can be testable with our channel, as mentioned above.
Conversely, models that modify only IBD without an accompanying ${}^{13}\mathrm C^\ast$ line are not constrained by this null search and will require complementary measurements (e.g., improved cross-sections and multi-channel comparisons).

Thorough analysis with our new channel is possible with the reduction of the dominant systematic uncertainty arising from the $\bar \nu_e {}^{13}{\rm C} \to \bar \nu_e {}^{13}{\rm C}^\ast$ cross-section precisions. 
To address this, we propose for the first time leveraging the capabilities of the current and future solar and accelerator neutrino experiments. 
The first and second categories mentioned above, i.e., a miscalculation of the flux and IBD systematics or the BSM contribution mimicking only the IBD, can be investigated 
with precise cross-section measurements in a wide range of energy region in various experiments such as JUNO, $\nu$EYE-IsoDAR at Yemilab, and $\pi$DAR or $\mu$DAR experiments such as OscSNS. 
We expect the future JUNO solar data, given the 1\% level ${}^8$B solar neutrino flux which can be obtained from measuring the oscillation parameters, can have the cross-section precision below 3\% with the background subtraction strategies discussed in Sec.~\ref{sec:xs} based on the series of the studies by the JUNO collaboration.
On the other hand, OscSNS and IsoDAR@Yemilab can have complementary measurements with the cross-section precision level below 4\% and 10\%, respectively.

In addition, we confirm that various current or near-future reactor experiments have potential to investigate the 5 MeV bump via our channel after implementing the background subtraction cuts suggested here, which are actually feasible in the case of JUNO-TAO based on the collaboration's studies.
At least 12 \% level flux separation capability between the observed 5 MeV bump and the theoretical models such as KI is required.
Equipped with the reduction of the systematic uncertainty from the cross-section measurements, JUNO-TAO can have sensitivity of 4\% which can even separate the theoretical flux models after 10 years of running.
PROSPECT-II can have sensitivity of 5\% level after giving up imposing $\beta/\gamma$ separation ability.
On the other hand, RENE which is essentially the follow-up program of NEOS II and RENO with the enhanced $\gamma$ capture ability, can have sensitivity below 8\%, separating the 5 MeV bump expectation and the HM model in our ${}^{13}$C channel. 
Extra background mitigation discussed in Appendix~\ref{appendix:moresubtraction} would significantly increase the sensitivities of those reactor experiments.
We also provide the method of fission fraction evolution in identifying the origin of the bump, which can help investigating the role of each isotope in contributing the neutrino spectrum around 5 MeV.

Our channel of inelastic neutrino - ${}^{13}$C scattering with a sharp 3.685 MeV photon emission can be also applicable to probe neutrinos from other sources such as the Sun, accelerator, and supernova.
We encourage the experimental colleagues to consider our proposal seriously and find a new avenue in clarifying the reactor neutrino flux in high energy region beyond 4 MeV.

\section*{Acknowledgments}

This work is supported by the Basic
Research Laboratory Program of the National Research Foundation (NRF) of Korea with Grant No. RS-2022-NR070815.
The work of PB, MGP, MR, and SS are additionally supported by the NRF with Grant No. RS-2020-NR055094 and RS-2025-00562917.
CSS is also supported by the NRF with Grant No. NRF-2022R1C1C1011840.
CSS and SS are partly supported by IBS-R018-D1.
The authors acknowledge the hospitality of APCTP during the focus program ``Dark Matter as a Portal to New Physics" (APCTP-2025-F01).

\nocite{*}

\appendix

\section{Contribution of the sterile neutrino non-standard interaction}
\label{appendix:nus}

In this section, we describe how we obtain the expected signals of 3.685 MeV photon from the model in Ref.~\cite{Berryman:2018jxt}. 
The scenario considers a sterile neutrino scattering with \(^{13}\text{C}\), producing a neutron and \(^{12}\text{C}^\ast\) in its first excited state ($\nu_s +{}^{13}{\rm C} \to \nu_s + n+ {}^{12}{\rm C}^\ast$) via the exchange of a new light gauge boson $X$ from the Lagrangian:
\begin{align}
\mathcal{L} = g_X X_\mu \left( Y_\nu \, \overline{\nu}_s \gamma^\mu \nu_s + \overline{p} \gamma^\mu p + Y_n \, \overline{n} \gamma^\mu n \right)\,,
\end{align}
where $g_X$ is the coupling constant of the new gauge symmetry U(1)$_X$ with $Y_\nu$ and $Y_n$ being the new gauge charges of the sterile neutrino and neutron, respectively, in units of $g_X$. 
The \(^{12}\text{C}^\ast\) subsequently decays into the ground state emitting a photon through \(^{12}\text{C}^\ast \to ^{12}\text{C} + \gamma\,(4.4\,\mathrm{MeV})\), mimicking the IBD signal together with the produced neutron~\cite{Berryman:2018jxt}.
Interestingly, this sterile neutrino scattering with ${}^{13}$C can also produce our signal: $\nu_s + {}^{13}{\rm C} \to \nu_s + {}^{13}C^*$ followed by the de-excitation ${}^{13}{\rm C}^\ast \to {}^{13}{\rm C} + \gamma\,(3.685\,{\rm \mathrm{MeV}})$ process.

We obtain the preferred parameter space explaining the 5 MeV bump in the RENO and NEOS experiments~\cite{RENO:2015ksa,NEOS:2016wee} by the sterile neutrino scenario producing $n + {}^{12}{\rm C}^\ast$ following the rough estimations in Ref.~\cite{Berryman:2018jxt}. 
The cross-section of the sterile neutrino scattering with ${}^{13}$C is expressed as
\begin{equation}
    \sigma_{\rm NP} = 4\pi  A_0(E) \zeta^2 \sigma_{\rm Mott}\,,   
\end{equation}
where \( A_0(E) \) is an experimentally determined function of the energy transferred to the nuclear system. 
The Mott cross-section, \( \sigma_{\rm Mott} \), is given by
\begin{equation}
\label{eq:Mott}
\sigma_{\rm Mott} = \frac{\pi \alpha^2 Z^2}{E_i E_f} \left[ \log \left( \frac{4 E_i E_f + M_X^2}{M_X^2} \right) - \frac{4 E_i E_f}{4 E_i E_f + M_X^2} \right],
\end{equation}
where \(E_i\) is the initial-state electron energy, \( E_f = E_i - E_n - E_{\rm th} \) is the final-state electron energy, and \( E_{\rm th} \) is the reaction threshold energy. 
The term  \( \zeta \)  is defined as
\begin{equation}
\label{eq:u1b}
\zeta =  Y_\nu \left(1 +\frac{7}{6}Y_n\right) \Big(\frac{g_X^2}{e^2}\Big)
 \sin \left( \frac{\Delta m_{41}^2 L}{4 E_\nu} \right) \sin 2\theta_{ee}\,,
\end{equation}
where \( \Delta m_{41}^2 \) is the new mass-squared splitting and \( \theta_{ee} \) is the new mixing angle. 
The U(1)$_X$ couplings for the 5 MeV bump imposed in Ref.~\cite{Berryman:2018jxt} are \( Y_\nu = +10 \), \( Y_n = -0.65 \), and \( \sin^2 2\theta_{ee} = 0.04 \).  
The oscillation part can be taken to be the averaged value 
\( \sin^2 \left( \frac{\Delta m_{41}^2 L}{4 E_\nu} \right   ) \to \frac{1}{2} \).

The parameter region of the U(1)$_X$ gauge boson mass, $M_X$, and the coupling $g_X$ explaining the 5 MeV bump observed in the RENO and NEOS experiments~\cite{RENO:2015ksa,NEOS:2016wee}
with 99\% C.L. are shown in Fig.~\ref{fig:sterile}.
Our analysis is consistent with the results of Ref.~\cite{Berryman:2018jxt}  
targeted 
at Daya Bay with 99\% confidence.   
Since this scenario partially can be tested from the COHERENT experiment, the current limit is shown with the black sold line following Ref.~\cite{Berryman:2018jxt} in the figure, strongly constraining the parameter region for $M_X \gtrsim 0.5$ MeV and $g_X \gtrsim 10^{-3}$.

As stated previously, the exactly same sterile neutrino scattering with ${}^{13}$C can also produce our signal, ${}^{13}{\rm C}^\ast \to {}^{13}{\rm C} + \gamma\,(3.685\,{\rm \mathrm{MeV}})$, which is expected to leave a sharp 3.685 MeV peak in the photon data in the reactor experiments.
In the first-order approximation, the cross-section 
of the sterile neutrino scattering with ${}^{13}$C producing the 3.685 MeV photon
can be expressed by replacing the Fermi constant, \( G_F \), with
\begin{equation}
    \frac{\zeta^2}{(p^2 - m_X^2) + i m_X\Gamma_X}\,,
    \label{eq:newcoup}
\end{equation}
where \( p \approx -Q = -3.685~\mathrm{MeV} \) and \( \Gamma_X = \frac{g_X^2 m_X}{24\pi} \). 
This contribution is at least thousand times larger than $G_F$.

Applying the current NEOS data collected for 180 days~\cite{Siyeon:2017tsg} and considering only the muon veto without discrimination of photons from other particles (such as electrons, positrons, and neutrons)
in the energy window of \( 3.685 \pm 0.1 \, \mathrm{MeV} \), the null observation of such a bump in NEOS provides the 99\% current upper bound on this scenario as the solid green curve in Fig.~\ref{fig:sterile}. This result completely excludes the 
possibility of the new physics scenario explaining the 5 MeV bump proposed in Ref.~\cite{Berryman:2018jxt}
with more than 99\% confidence 
even with the current data.
This is due to the fact that the sterile neutrino NSI contribution to the inelastic $\bar \nu_e$ - ${}^{13}$C scattering is at least million times larger than the SM process via $Z$ boson exchange, due to the difference of $G_F$ and Eq.~\ref{eq:newcoup}.
Assuming 10 years of data collection without particle discrimination, the 99\% upper bound is represented by the 
dot-dashed curve. 
It is possible to substantially enhance the capability of NEOS in probing new physics scenarios by reducing the backgrounds to the level comparable to the number of signals following the background subtraction strategies explained in Sec.~\ref{sec:xs} and Appendix~\ref{appendix:moresubtraction} (for further mitigation).
Those sensitivities are illustrated as the dashed and dotted curves representing 180 days and 10 years of data collection, respectively.

\section{Further background mitigation strategy}
\label{appendix:moresubtraction}

This section is devoted to our proposal of mitigating the backgrounds further in order to effectively use the neutrino -${}^{13}$C scattering process emitting the 3.685 MeV photon as an alternative channel in detecting reactor neutrinos.
Basically our proposal pursues a minimum modification such as having extra shielding and moderators to already existing detectors.
For a detector close to the reactor such as NEOS and its upgrade version RENE~\cite{Choi:2025wbw}, the background sources are dominated by neutron- and muon-induced events. 
Additional backgrounds include misidentification of electrons from $\nu$ES or IBD events caused by reactor neutrinos, which scale with the signal events. Other contributions come from thallium ($^{208}$Tl) $\beta$ decays due to thorium ($^{232}$Th) contamination in the LS detector, as well as environmental backgrounds originating from external radioactivity. The contribution of solar neutrinos is negligible for short-baseline reactor experiments but becomes significant if $\left(\text{Baseline/km}\right)^2 \gg \left(\text{Power/GW}\right)$.

Since the reactor neutrinos are currently detected via IBD, moderate neutron shielding and muon veto systems are sufficient for basic background reduction. However, to effectively separate the background from the \(^{13}\text{C}\) signal, more advanced background reduction techniques are needed, particularly for distinguishing between flux models or generally improving sensitivity. We demonstrate the feasibility of our proposal by implementing additional shielding in the currently installed one-ton RENE detector, located 24 meters from a 2.8 GW Hanbit reactor\cite{yang2024rene}.

Assuming an energy resolution of $5\% / \sqrt{\text{E (MeV)}}$, we define our ROI as 3.685 MeV $\pm$ the full width at half maximum (FWHM) of 0.1 MeV. For comparison, the energy resolution of JUNO-TAO is $1\% / \sqrt{\text{E (MeV)}}$~\cite{JUNO:2020ijm}. 
The current RENE detector incorporates a gamma catcher to reduce gamma background and enhance energy resolution. 
The shielding for RENE consists of multiple layers: 46 cm of water, 40 cm of borated polyethylene (BPE) and high-density polyethylene (HDPE) on top, 17.5 cm of BPE and HDPE on the sides, and 7 cm of steel, 2.5 cm of lead, and 2.5 cm of aluminum. 
Based on simulation results~\cite{lee2024hfir,heffron2017characterization,hackett2017dang,norcini2019first,ashenfelter2016background,kyzylova2021characterization}, we estimate approximately $10^6$ thermal neutron events in a one-ton detector within the deposit energy range of $3.685 \pm 0.1~\mathrm{MeV}$ over one year. In the Tendon gallery, where NEOS is located and RENE will be installed, the fast neutron background is expected to result in approximately $10^4$ events per year~\cite{NEOS:2016kan}. With a 99\% photon-neutron discrimination efficiency, the number of fast neutron-induced background events is expected to be reduced to around 100 per year.

In order to reduce the thermal neutron induced backgrounds, extra shielding such as water, polyethylene, borated polyethylene, and cadmium in addition to the current RENE. 
Photon-neutron discrimination efficiency in the LS detectors typically ranges from 90\% to 99\%, depending on factors such as the specific experiment, energy range, and optimization of the pulse shape discrimination (PSD) method. Notice that the discrimination efficiency is generally higher for fast neutrons than for thermal neutrons~\cite{NEOS-II:2022mov}.
For thermal neutrons, by adding an extra half-meter of water, with a macroscopic removal cross-section for thermal neutrons of $\Sigma_r = 0.17~\text{cm}^{-1}$, its flux can be reduced by a factor of $10^{-4}$ (since $e^{-8.5} \approx 2 \times 10^{-4}$). 
This brings the number of thermal neutron events down to approximately 100 per year.  
Considering 90\% to 99\% photon-neutron discrimination is achievable, 
the expected thermal neutron background is of the order of ten to one event per year.

To reduce the fast neutron backgrounds further, moderators can be employed to slow them down to thermal energies. Hydrogen-rich materials, such as water, polyethylene, and paraffin, effectively slow fast neutrons through elastic scattering, as hydrogen nuclei have a mass similar to that of neutrons, enabling efficient energy transfer during collisions. Once slowed, fast neutrons can then be captured or absorbed.
Neutron absorbers such as boron or cadmium are effective at capturing thermal neutrons post-moderation. High-$Z$ materials like lead, steel, and tungsten are excellent for capturing fast neutrons and also provide additional shielding against secondary gamma radiation. 
Overall, while thermal neutrons can be reduced using shielding and boron-doped materials, fast neutron rates can be managed and monitored through the use of extra shielding, and photon-neutron discrimination techniques.

For very short-baseline detectors with an overburden of a few tens of meters water equivalent (m.w.e.), between $10^2$ and $10^3$ muons pass through each square meter of the detector per second. In contrast, at sites like the RENO near detector, Daya Bay, and Double Chooz, with overburdens of several hundred m.w.e., the muon rates are significantly lower, at less than one muon per second per square meter. The current shielding for the RENE detector reduces the muon background by 60\%.
For NEOS, the total number of muon events in a one-ton detector is approximately $10^7$ events per year within the energy range of $3.685 \pm 0.1$ MeV. Applying the muon cut at NEOS eliminates 95\% of these events within the specified energy range. However, using advanced active veto systems and discrimination techniques, such as those employed by Borexino, more than 99.99\% of muons can be effectively rejected~\cite{Borexino:2013cke}. In the Borexino experiment, the cosmogenic muon background is mitigated using a combination of an active veto system and muon-electron discrimination based on scintillation and Cherenkov light. However, since the primary 3.685 MeV photon does not directly produce Cherenkov light (unless secondary charged particles are involved), better discrimination between cosmogenic muon-induced backgrounds and 3.685 MeV photon signals is expected in principle. Additionally, applying a magnetic field can deflect muons, further improving their discrimination from photons.
By enhancing the energy resolution of the detector, improving shielding, implementing an effective muon veto system, and employing advanced photon-muon discrimination techniques and/or a magnetic field, it is possible to reduce muon background events to the level of the signal or lower, even for very short-baseline detectors with minimal overburden. 

For detectors with significant overburden, such as near and far detector halls of RENO, Daya Bay, and Double Chooz, only a few muon background events are expected per kton.year. Moreover, these experiments with larger baselines, neutron-induced backgrounds are less significant. Instead, ES becomes the dominant source of background. 
Adopting the neutron veto capability along with a fiducial volume cut in Ref.~\cite{Conrad:2004gw}, we expect the misidentified ES events is about 6 times the signal events. 
For JUNO-TAO with a good energy resolution and the $\beta/\gamma$ discrimination efficiency about 95\% that we expect to achievable in our ROI, the number of ES and IBD misidentified background is expected to be 15 times smaller than our signal~\cite{JUNO:2020ijm}. 
Another source of background arises from misidentified IBD events, which constitute approximately 0.1\% of the ES events.

Additionally, a significant source of background below 5 MeV comes from the beta decays of \(^{208}\text{Tl}\) in the \(^{232}\text{Th}\) decay chain, accompanied by simultaneous emissions of \(\alpha\) and \(\beta\), as stated earlier. 
For a LS detector with \(^{232}\text{Th}\) contamination at the level of \(5\times10^{-17}\,\text{g/g}\) (equivalent to KamLAND-level purity), we expect around one \(^{208}\text{Tl}\) event within the ROI per day in a 1 kt LS detector. 
However, by exploiting the coincidence between the \(\alpha\)-decay in the \(^{232}\text{Th}\) chain and the \(\beta\)-decay of \(^{208}\text{Tl}\)—a technique known as \(^{232}\text{Th}\)-series tagging—80\% and 99\% of this background can be rejected in KamLAND and JUNO, respectively~\cite{Hachiya:2020mpu,JUNO:2020hqc}. Even under a conservative assumption of 80\% background rejection, the \(^{232}\text{Th}\)-related background would remain subdominant compared to the ES events.

Environmental backgrounds such as external radio-activities from detector walls and surrounding rocks also exist.
However, fiducial volume cut can reduce those backgrounds to the 5\% and negligible level, respectively~\cite{Conrad:2004gw}.
These environmental backgrounds can be further mitigated using reactor on-off time and the  $\beta/\gamma$ discrimination.

In summary, the primary source of background for very short baseline experiments is cosmogenic muons and neutron-induced backgrounds, whereas for longer baseline experiments, such as the near and far halls of Daya Bay and RENO, the dominant background arises from electron scattering misidentification. 
We expect the former backgrounds can be enormously mitigated by adding the extra shielding and moderators, photon-neutron discrimination techniques, and muon veto systems to the RENE detector.
We encourage the experimental colleagues to consider our proposal and probe reactor antineutrinos via the multiple detection channels.

\bibliographystyle{apsrev4-1}

\bibliography{5mev}

\end{document}